\documentclass[aps,prl,groupedaddress,showpacs,showkeys]{revtex4-1}
\usepackage{graphicx}% Include figure files
\usepackage{dcolumn}% Align table columns on decimal point
\usepackage{bm}% bold math
\def\mbi#1{\mbox{\bfseries\itshape #1}} 

\begin{document}

%\preprint{}

\title{Effects of Primordial Magnetic Field with Log-normal Distribution on Cosmic Microwave Background}


\author{Dai G. Yamazaki$^{1}$}
% \homepage{http://th.nao.ac.jp/~yamazaki/}
 \email{yamazaki.dai@nao.ac.jp}
\author{Kiyotomo Ichiki$^{2}$}%
\author{Keitaro Takahashi$^{3}$}%
%\thanks{}
\affiliation{% 
$^{1}$
Division of Theoretical Astronomy, National Astronomical Observatory of Japan,
2-21-1, Osawa, Mitaka, Tokyo, 181-8588, Japan 
}%
\affiliation{% 
$^{2}$Department of Physics and Astrophysics, Nagoya University, 
Furo-cho, Chikusa-ku, Nagoya, Aichi, 464-8602, Japan
}%
\affiliation{% 
$^{3}$
Faculty of Science, Kumamoto University, 
2-39-1, Kurokami, Kumamoto, Kumamoto, 860-8555, Japan
}%


\date{\today}

\begin{abstract}
We study the effect of primordial magnetic fields (PMFs) on the anisotropies
of the cosmic microwave background (CMB). We assume the spectrum of PMFs
is described by log-normal distribution which has a characteristic scale,
rather than power-law spectrum. This scale is expected to reflect
the generation mechanisms and our analysis is complementary to previous
studies with power-law spectrum. We calculate power spectra of energy density
and Lorentz force of the log-normal PMFs, and then calculate CMB
temperature and polarization angular power spectra from scalar, vector,
and tensor modes of perturbations generated from such PMFs. By comparing
these spectra with WMAP7, QUaD, CBI, Boomerang, and ACBAR data sets,
we find that the current CMB data set places the strongest constraint at
$k\simeq 10^{-2.5}$ Mpc$^{-1}$ with the upper limit $B\lesssim 3$ nG.
\end{abstract}
%\keywords{}
%must follow title, authors, abstract, \pacs, and \keywords
%Use showkeys c
\maketitle 
%_/_/_/_/_/_/_/_/_/_/_/_/_/_/_/_/_/_/_/_/_/_/_/_/_/
%_/_/_/_/_/_/_/_/_/_/_/_/_/_/_/_/_/_/_/_/_/_/_/_/_/
%_/_/_/_/_/_/_/_/_/_/_/_/_/_/_/_/_/_/_/_/_/_/_/_/_/
\section{\label{sec:introduction}Introduction}
%_/_/_/_/_/_/_/_/_/_/_/_/_/_/_/_/_/_/_/_/_/_/_/_/_/
%_/_/_/_/_/_/_/_/_/_/_/_/_/_/_/_/_/_/_/_/_/_/_/_/_/
%_/_/_/_/_/_/_/_/_/_/_/_/_/_/_/_/_/_/_/_/_/_/_/_/_/
Many researchers have studied magnetic fields for wider ranges
in the early Universe. Recently, magnetic fields with strength
$\sim 1 \mu$ G have been observed in clusters of galaxies, and
primordial magnetic fields (PMFs) have been studied by many authors
to explain their origins. The strength of PMFs is constrained as
$B\lesssim 1$ nG from cosmological observations, such as temperature
and polarization anisotropies of cosmic microwave background
(CMB) and matter power spectra
\cite{Subramanian:1998fn, Mack:2001gc, Subramanian:2002nh,
Lewis:2004ef, Yamazaki:2004vq, Kahniashvili:2005xe, Challinor:2005ye,
Dolgov:2005ti, Gopal:2005sg, Yamazaki:2005yd, Kahniashvili:2006hy,
Yamazaki:2006mi, Yamazaki:2006bq, Yamazaki:2006ah, Giovannini:2006kc,
Yamazaki:2007oc, Paoletti:2008ck, Finelli:2008xh, Yamazaki:2008bb,
2008nuco.confE.239Y, Sethi:2008eq, Kojima:2008rf, 2008PhRvD..78f3012K,
Giovannini:2008aa, Yamazaki:2009na, Shaw:2009nf, 2010PhRvD..81b3008Y,
2010PhRvD..81j3519Y, Yamazaki2010aa}.

Many theoretical models have been proposed for generating PMFs of
cosmological scales. Generation models of PMFs from inflation can 
create strong fields, whose amplitude is of the order of $1$ nG at 
redshift $z\sim 0$, depending on assumed hypothetical fields and couplings
\cite{Turner:1987bw,Ratra:1991bn,Bamba:2004cu,2004PhRvD..69d3507B,
2007JCAP...02..030B}. 
Furthermore, PMFs can be generated by primordial perturbations of
density fields \cite{Takahashi:2005nd,ichiki:2006sc}, the Beiermann battery
mechanism in primordial supernovae remnants \cite{Hanayama:2005hd}
or the Weibel instabilities \cite{2005MNRAS.364..247F}.
These PMFs can evolve into the magnetic fields observed in
galaxies and/or galaxy clusters directly or through the dynamo process.

While magnetogenesis during inflation can produce PMFs beyond
the horizon scale, PMFs generated by the other mechanism are expected
to have the characteristic scale because they are based on causal
processes. Here it should be noted that the coherence length of
PMFs can grow after generation due to inverse cascade
\cite{2001PhRvE..64e6405C,Banerjee:2004df}, although the efficiency is still
under study.

In the previous studies to put constraints on PMFs from CMB anisotropies,
the spectrum of PMFs was assumed to be power-law shape. 
Although power-law spectrum is natural for magnetogenesis during inflation,
this is not the case for the other mechanisms based on causal processes.
Furthermore, in previous studies with power-law PMFs, it was not so clear
which scale of PMFs mainly contributes to the constraints.
Since inflationary mechanisms tend to generate power-law magnetic fields, the
magnetic fields which have a characteristic scale at cosmological scales are
difficult to be produced and seem somewhat artificial. However, it is still
possible that such fields can be constrained by observation, and if observed it
would give us useful information about generation mechanism of cosmological
magnetic fields.  Hence, in this article, as a toy example we use a log-normal
distribution (LND) for the PMF spectrum
$f_\mathrm{LND}(k;k_\mathrm{M},\sigma_\mathrm{M})$ which has a characteristic
scale expressed by $k_\mathrm{M}$, and aim to constrain PMFs scale by scale
instead of the power-law PMF spectrum.
These quantities reflect
the generation mechanism but here we regard them just as parameters.
We study features of angular power spectra of the CMB from the PMFs
with log-normal distribution. Finally we constrain the strength
of the PMFs for fixed sets of the parameters.
%_/_/_/_/_/_/_/_/_/_/_/_/_/_/_/_/_/_/_/_/_/_/_/_/_/_/_/_/_/_/_/_/_/_/_/
%_/_/_/_/_/_/_/_/_/_/_/_/_/_/_/_/_/_/_/_/_/_/_/_/_/_/_/_/_/_/_/_/_/_/_/
\section{Power Spectrum of PMF}
%_/_/_/_/_/_/_/_/_/_/_/_/_/_/_/_/_/_/_/_/_/_/_/_/_/_/_/_/_/_/_/_/_/_/_/
%_/_/_/_/_/_/_/_/_/_/_/_/_/_/_/_/_/_/_/_/_/_/_/_/_/_/_/_/_/_/_/_/_/_/_/
The electromagnetic tensor has the usual form
\begin{eqnarray}
{F^\alpha}_\beta=
\left(
    \begin{array}{cccc}
      0    &  E_1  &  E_2  &  E_3 \\
      E_1  &  0    & -B_3  &  B_2 \\
      E_2  &  B_3  &  0    & -B_1 \\
      E_3  & -B_2  &  B_1  &  0  \\
    \end{array}
\right)~~,
\label{eq_emf_tensor}
\end{eqnarray}
where $E_i$ and $B_i$ are the electric and magnetic fields.
Here we use natural units $c = \hbar = 1$. 
The energy-momentum tensor for electromagnetism is 
\begin{eqnarray}
{T^{\alpha\beta}}_{[\mathrm{EM}]}=\frac{1}{4\pi}\left(F^{\alpha\gamma}F^\beta_\gamma
-\frac{1}{4}g^{\alpha\beta}F_{\gamma\delta}F^{\gamma\delta}\right)\label{eq_ememtensor}.
\end{eqnarray}
The Maxwell stress tensor, $\sigma^{i k}$ is derived from the space-space components of the electromagnetic energy-momentum tensor,  
\begin{eqnarray}
-{T^{ik}}_{[\mathrm{EM}]}=
\sigma^{ik}=  
\frac{1}{4\pi a^2}\left\{E^i E^k+B^i B^k
- \frac{1}{2}\delta^{ik}(E^2+B^2)\right\}\label{eq_MST1}.
\end{eqnarray}
where $a$ is the cosmic scale factor.

We consider the case that the spectrum of the PMFs at large scales
had been well established before the CMB epoch, after their generation and 
followed by perhaps a complicated time evolution. In this case we do not need
to consider the time evolution of magnetic fields in our CMB analysis,
e.g. nonliner effects \cite{Banerjee:2004df}, of the PMF and we consider
only the dilution due to the expansion of the Universe.
Within the linear approximation \cite{Durrer:1999bk} we can discard the MHD
backreaction onto the field itself. The conductivity of the primordial
plasma is very large, and magnetic fields are frozen in the plasma
\cite{Mack:2001gc}. This is a very good approximation during  the epochs
of interest here. Furthermore, we can neglect the electric field,
i.e.~$E\sim 0$, and can decouple the time evolution of the magnetic
field from its spatial dependence, i.e.
$\mathbf{B}(\tau,\mathbf{x}) = \mathbf{B}(\mathbf{x})/a^2$ for very
large scales. In this way we obtain the following equations,
\begin{eqnarray}
{T^{00}}_{[\mathrm{EM}]}=\frac{B^2}{8\pi a^6} \label{eq_MST_00}~~, \\
{T^{i0}}_{[\mathrm{EM}]}={T^{0k}}_{[\mathrm{EM}]}=0 \label{eq_MST_0s} ~~,\\
-{T^{ik}}_{[\mathrm{EM}]}=\sigma^{ik}=\frac{1}{8\pi a^6}(2B^i B^k -
\delta^{ik}B^2)~~.
\label{eq_MST_ss}
\end{eqnarray}
%_/_/_/_/_/_/_/_/_/_/_/_/_/_/_/_/_/_/_/_/_/_/_/_/_/
%_/_/_/_/_/_/_/_/_/_/_/_/_/_/_/_/_/_/_/_/_/_/_/_/_/
\subsection{power spectra of PMFs from the log-normal distribution}
%_/_/_/_/_/_/_/_/_/_/_/_/_/_/_/_/_/_/_/_/_/_/_/_/_/
%_/_/_/_/_/_/_/_/_/_/_/_/_/_/_/_/_/_/_/_/_/_/_/_/_/
We defined a two point correlation function of the PMFs as follows,
\begin{eqnarray}
\left\langle 
	B^{i}(\mbi{k}) {B^{j}}^*(\mbi{k}')
\right\rangle 
	&=& 
	(2\pi)^3P_\mathrm{[PMF]}(k)P^{ij}(k)\delta(\mbi{k}-\mbi{k}')~,
\label{eq:two_point_correlations_of_PMF}
\end{eqnarray}
where
\begin{eqnarray}
P^{ij}(k)&=&
	\delta^{ij}-\frac{k{}^{i}k{}^{j}}{k{}^2}.\label{project_tensor}
\end{eqnarray}
We follow the convention for the Fourier transform as
\begin{eqnarray}
f(\mbi{k}) = \int d^3 x \exp (i\mbi{k} \cdot \mbi{x}) F(\mbi{x}).
\end{eqnarray}

If PMFs have been generated from inflation the power-law model would
be a good representation of the magnetic field power spectrum. 
On the other hand, if the PMFs were created through mechanisms other than
inflation, the spectrum would have a characteristic scale and may not
be described as a simple power-law.

Furthermore, we need to understand how such PMFs cascade
from smaller to larger scales to study effects of the PMFs created in
 the early Universe on the large-scale structures.  
Actually, we can learn behaviors of PMF cascading or inverse cascading by simulations\cite{2010PhRvD..81l3002K}, whose results, however, generally depend on
cosmological parameters and models. 
Since such simulations generally take too much time
 and have a limited dynamical range, it is not efficient to
estimate distributions of the PMF by some simulations when we study
cosmology and astrophysics with the PMF quantitatively.

To avoid these problems, we use in place of the power law a LND defined as,
\begin{eqnarray}
f_\mathrm{LND}(k;k_\mathrm{M}, \sigma_\mathrm{M})
=
\frac{1}{k \sigma_\mathrm{M} \sqrt{2\pi}}
	\exp \left\{
			-\frac{
				\left[
					\ln{(k)}-\ln{(k_\mathrm{M})}
				\right]^2
				}
				{2\sigma_\mathrm{M}^2}
		 \right\}
\label{eq:power_lg},
\end{eqnarray}
where $k_\mathrm{M}$ is the characteristic scale depending on the PMF
generation model, and $\sigma_\mathrm{M}$ is the variance.  The
variance $\sigma_\mathrm{M}$ in Eq.(\ref{eq:power_lg}) expresses how
this distribution is expanded (or concentrated) around the
characteristic scale $k_\mathrm{M}$.  
Therefore, the variance $\sigma_\mathrm{M}$ may be related to the cascading of the PMFs.
{Using Eq.(\ref{eq:power_lg}), the power spectrum of the PMF is given by}
\begin{eqnarray}
  P_\mathrm{[PMF]}(k) = A f_\mathrm{[LND]}(k;k_\mathrm{M},\sigma_\mathrm{M}).
\end{eqnarray}

We shall determine the coefficient from the variance of magnetic fields in real space,
\begin{eqnarray}
\left.
	\left\langle
		B^{i}(\mbi{x}) {B^{i}}(\mbi{x})
	\right\rangle
\right|_d
= B_d^2
\label{eq:real}
\end{eqnarray}
where $d = 2 \pi/ k_\mathrm{M}$ and
\begin{eqnarray}
\left\langle
	B^{i}(\mbi{x}) {B^{j\ast}}(\mbi{x}')
\right\rangle
=
	\frac{1}{(2\pi)^6}
	\int d^3 k 
	\int d^3 k'
		\exp(-i\mbi{x}\cdot\mbi{k}+i\mbi{x}'\cdot\mbi{k}') 
		\left\langle
			B^{i}(\mbi{k}) {B^{j\ast}}(\mbi{k}')
			\times w(k)
		\right\rangle.\label{eq:real2}
\end{eqnarray}
 The window function $w(k)$ is given by the rectangular function
as $w(k)=1 $ for $ k_-< k <k_+$ and $w(k)=0 $ for otherwise.
Here,  
\begin{eqnarray}
\left\{
\begin{array}{l}
k_{+} = \exp{(\ln{k_\mathrm{M}}+n\sigma_\mathrm{M})}={10^{\frac{\ln{k_\mathrm{M}}+n\sigma_\mathrm{M}}{\ln{(10)}}}},\\
k_{-} = \exp{(\ln{k_\mathrm{M}}-n\sigma_\mathrm{M})}={10^{\frac{\ln{k_\mathrm{M}}-n\sigma_\mathrm{M}}{\ln{(10)}}}},
\end{array}
\right.
\end{eqnarray}
where $n$ gives the range of integration.
From Eq.(\ref{eq:real2}), we obtain the amplitude of the log-normal distribution as
\begin{eqnarray}
A = B^2_\mathrm{d}
	\frac{(2\pi)^2}{4}
	\left\{
		\int^{k_{+}}_{k_{-}} dk 
		k
		\frac{1}{\sigma_\mathrm{M}\sqrt{2\pi}}
		\exp{
		\left[
			-\frac{(\ln |k|-\ln |k_\mathrm{M}|)^2}{2\sigma_\mathrm{M}^2}
		\right]
		}
	\right\}^{-1}.
\label{eq:rect_A}
\end{eqnarray}
As mentioned above, we can
neglect the diffusion of the magnetic field at cosmological scales for
the age of the Universe. Therefore we can assume that the energy of the
LND-PMF is not dissipated; in other words, the integral of Eq.(11) over
$k$ from $k_-$ to $k_+$ is time-independent. In this case, $k_-$ and
$k_+$ should be taken to be $\pm$ infinites. In practice,
we take $k_-$ and $k_+$ far enough from the peak of the distribution
\footnote{we take $n= 6$, in the present study, and the integral of
Eq.(10) over $k$ form $k_-$ to $k_+$ is 0.999999998027c}.

Using Eq.(\ref{eq:power_lg}) and the
method presented by 
Ref.\cite{Yamazaki:2006mi,Yamazaki:2007oc}, we obtain power spectra from
the LND function for each perturbation mode as follows.
For the scalar mode, the power
spectrum of the energy (pressure) of the PMFs is
\begin{eqnarray}
P_\mathrm{[EM:SE]}(k;k_\mathrm{M}, \sigma_\mathrm{M})=
&&
\frac{A^2}{8(2\pi)^4 a^8}
\int dk'{k'}^{2}\int^{1}_{-1} d\mathcal{C}
	f_\mathrm{LND}(|k'|;k_\mathrm{M},\sigma_\mathrm{M})
	f_\mathrm{M}(|k-k'|;k_\mathrm{M},\sigma_\mathrm{M})
\nonumber\\
&&
\times
\frac{
\left\{
(1+\mathcal{C}^2)k^2-4kk'\mathcal{C}+2{k'}^2
\right\}
}
{
|\mbi{k}-\mbi{k}'|^{2}
}~,
\nonumber\\
\label{p:se}
\end{eqnarray}
the power spectrum of the magnetic tension is
\begin{eqnarray}
P_\mathrm{[EM:ST]}(k;k_\mathrm{M}, \sigma_\mathrm{M})=
&&
\frac{A^2}{2(2\pi)^4 a^8}
	\int
		dk'k'{}^2
	\int^{1}_{-1} d\mathcal{C}
	f_\mathrm{M}(|k'|;k_\mathrm{M},\sigma_\mathrm{M})
	f_\mathrm{M}(|k-k'|;k_\mathrm{M},\sigma_\mathrm{M})
\nonumber\\
&&
\times
		\frac{k'{}^2}
		{|\mbi{k}-\mbi{k}'|^2}
		\left(
			1-\mathcal{C}^2
		\right)^2~,
\nonumber\\
\label{p:st}
\end{eqnarray}
and the correlation term between the energy and tension is
\begin{eqnarray}
P_\mathrm{[EM:SET]}(k;k_\mathrm{M}, \sigma_\mathrm{M})=
&&
\frac{A^2}{4(2\pi)^4 a^8}
	\int dk' {k'}^2
	\int^{1}_{-1} d\mathcal{C}
	f_\mathrm{LND}(|k'|;k_\mathrm{M},\sigma_\mathrm{M})
	f_\mathrm{LND}(|k-k'|;k_\mathrm{M},\sigma_\mathrm{M})
\nonumber\\
&&
\times
	\frac{k'(1-\mathcal{C}^2)}
		 {|\mbi{k}-\mbi{k}'|^2}
	(k'-k\mathcal{C})~.
\nonumber\\ 
\label{p:set}
\end{eqnarray}
For the vector mode, the power spectrum of the PMFs
is;
\begin{eqnarray}
P_\mathrm{[EM:V]}(k;k_\mathrm{M}, \sigma_\mathrm{M})=
&&
\frac{A^2}{8 (2\pi)^4 a^8}
	\int
		dk'
		k'{}^2
\int^{1}_{-1} d\mathcal{C}
	f_\mathrm{LND}(|k'|;k_\mathrm{M},\sigma_\mathrm{M})
	f_\mathrm{LND}(|k-k'|;k_\mathrm{M},\sigma_\mathrm{M})
\nonumber\\
&&
\times
  \frac{(1-C^2)}{|\mbi{k}-\mbi{k}'|^2}
  \left\{
    2k^2-5Ckk'
    +(2C^2+1)k'{}^2
  \right\}~.
\nonumber\\ 
\label{p:v}
\end{eqnarray}
Finally, for the tensor mode the power spectrum of
anisotropic stress of
the PMFs is found to be
\begin{eqnarray}
P_\mathrm{[EM:T]}(k;k_\mathrm{M}, \sigma_\mathrm{M})=
&&
	\frac{A^2}{16(2\pi)^4 a^8}
	\int
		dk'k'{}^2
	\int^{1}_{-1} d\mathcal{C}
	f_\mathrm{LND}(|k'|;k_\mathrm{M},\sigma_\mathrm{M})
	f_\mathrm{LND}(|k-k'|;k_\mathrm{M},\sigma_\mathrm{M})
\nonumber\\
&&
\times
		\{1+C^2\}\left\{1+\frac{(k-Ck')^2}{|\mbi{k}-\mbi{k}'|^2}\right\}~.
\nonumber\\ 
\label{p:t}
\end{eqnarray}
Since it is difficult to analytically make convolutions of the log-normal distribution functions, we numerically estimate them.
%_/_/_/_/_/_/_/_/_/_/_/_/_/_/_/_/_/_/_/_/_/_/_/_/_/
%_/_/_/_/_/_/_/_/_/_/_/_/_/_/_/_/_/_/_/_/_/_/_/_/_/
%_/_/_/_/_/_/_/_/_/_/_/_/_/_/_/_/_/_/_/_/_/_/_/_/_/
\section{Evolution Equations}
%_/_/_/_/_/_/_/_/_/_/_/_/_/_/_/_/_/_/_/_/_/_/_/_/_/
%_/_/_/_/_/_/_/_/_/_/_/_/_/_/_/_/_/_/_/_/_/_/_/_/_/
%_/_/_/_/_/_/_/_/_/_/_/_/_/_/_/_/_/_/_/_/_/_/_/_/_/
In this section, we shall summarize the essential evolution equations
for each mode. For details, see Refs.\cite{Yamazaki:2007oc,Shaw:2009nf}.
In this paper, we use the metric perturbations of the conformal Newtonian gauge defined by Ref.\cite{Ma:1995ey,Hu:1997hp,Shaw:2009nf}. 
We also express a scalar potential which is the gravitational potential in the Newtonian limit by $\psi$, and the other scalar potential by $\phi$. 
The vector and tensor potentials in the metric perturbations are expressed by $V$ and $\mathcal{H}$.
The subscripts "CDM", "b", "$\gamma$" and "$\nu$" in the equations in this paper indicate cold dark matter, baryon, photon and neutrino, respectively
\subsection{Boltzmann equations and Hierarchy}
We can understand the time evolution of temperature perturbations and polarization anisotropies of CMB photons from the Boltzmann equations.
In this subsection, we introduce the Boltzmann equations as follows (in detail see Ref. 
\cite{Hu:1997hp}),
\begin{eqnarray}
\dot{\mathcal{Q}}^{(M)}_{\ell} &=& 
	k
	\left[
		  \frac{\ell^2-M^2}{2\ell-1}
		  \mathcal{Q}^{(M)}_{\ell-1}
		- \frac{(\ell+1)^2-M^2}{2\ell+3}
		  \mathcal{Q}^{(M)}_{\ell+1}
	\right]
\nonumber\\
&& 
- a n_e \sigma_T \mathcal{Q}^{(M)}_{\ell}
+ \mathcal{S}^{(M)}_{\ell}, (\ell \ge M)
\end{eqnarray}
where $n_e$ is a number density of free electrons and 
$\sigma_T$ is the Thomson scattering cross section.
Here, an index of $M$ is a kind of perturbation mode, and 
$M=0, \pm 1, \pm, 2$ show scalar, vector and tensor modes respectively.
The third term $\mathcal{S}^{(M)}_{\ell}$ shows effects of gravitational potentials and scattering by other matters. 
We can write down nonzero terms of $\mathcal{S}^{(M)}_{\ell}$ as follows:
\begin{eqnarray}
\mathcal{S}^{(0)}_0  =
	\tau_C \mathcal{Q}^{(0)}_0 - \dot{\phi},~
\mathcal{S}^{(0)}_1  =
	\tau_C v^\mathrm{(S)}_b + k{\psi},~
\mathcal{S}^{(0)}_2  = \tau_C \dot{\mathcal{P}}^{(0)},
\\
\mathcal{S}^{(1)}_1  =
	\tau_C v^\mathrm{(V)}_b + \dot{V},~
\mathcal{S}^{(1)}_2  = \tau_C \dot{\mathcal{P}}^{(1)},
\\
\mathcal{S}^{(2)}_2  = \tau_C \dot{\mathcal{P}}^{(2)} - \dot{\mathcal{H}},
\end{eqnarray}
where $\tau_C = a n_e \sigma_T$ is the differential optical depth of
Compton scattering,
and $\mathcal{P}^{M}$ are the anisotropic effects from Compton scattering
and polarizations given by
\begin{eqnarray}
\mathcal{P}^{M} = 
	\frac{1}{10}
	\left[
		  \mathcal{Q}^{(M)}_2
		- \sqrt{6}\mathcal{E}^{(M)}_2
	\right]~.
\end{eqnarray}
Here, $\mathcal{E}^{(M)}_\ell$ and $\mathcal{B}^{(M)}_\ell$ 
are $E$ and $B$ mode polarizations,
respectively, whose evolutions are given by
\begin{eqnarray}
\dot{\mathcal{E}}^{(M)}_{\ell} &=& 
	k
	\left[
		  \frac{_2\epsilon^M_\ell}{2\ell-1}
		  \mathcal{E}^{(M)}_{\ell-1}
		- \frac{{2M}}{\ell(\ell+1)}
		  \mathcal{B}^{(M)}_{\ell+1}
		- \frac{_2\epsilon^M_{\ell+1}}{2\ell+3}
		  \mathcal{E}^{(M)}_{\ell+1}
	\right]
\nonumber\\
&& 
- \tau_C 
  \left[
	  \mathcal{E}^{(M)}_{\ell}
	+ \sqrt{6}\mathrm{P}^{(M)}\delta_{\ell,2}
  \right]~,
\end{eqnarray}
\begin{eqnarray}
\dot{\mathcal{B}}^{(M)}_{\ell} &=& 
	k
	\left[
		  \frac{_2\epsilon^M_\ell}{2\ell-1}
		  \mathcal{B}^{(M)}_{\ell-1}
		+ \frac{{2M}}{\ell(\ell+1)}
		  \mathcal{E}^{(M)}_{\ell+1}
		- \frac{_2\epsilon^M_{\ell+1}}{2\ell+3}
		  \mathcal{B}^{(M)}_{\ell+1}
	\right]
\nonumber\\
&& 
- \tau_C 
	  \mathcal{B}^{(M)}_{\ell},
\end{eqnarray}
where
\begin{eqnarray}
_h\epsilon^M_\ell &=& 
	\sqrt{
			\frac{(\ell^2-M^2)(\ell^2-h^2)}{\ell^2}
		}~.
\end{eqnarray}
\subsection{Scalar Mode}
From 
Refs.\cite{Padmanabhan:1993booka,
Ma:1995ey,
Hu:1997hp,
Hu:1997mn,
Dodelson:2003booka,
Giovannini:2006gz,
Yamazaki:2007oc,
Shaw:2009nf}, we obtain the same form for the evolution
 equations of photons and baryons as in previous
 work \cite{Padmanabhan:1993booka,Ma:1995ey,Hu:1997hp,Hu:1997mn,Dodelson:2003booka}, by considering the Compton interaction
 between baryons and photons,  
\begin{eqnarray}
k^2\phi + 3H(\dot{\phi}+H\psi) &=& 4\pi G{a^2}
\left\{
	E_\mathrm{[EM:S]}(\mbi{k},\tau)-\delta\rho_\mathrm{tot}
\right\} \label{eq:phi}\\
k^2(\phi-\psi) &=& 
-12\pi G{a^2}
\left\{
	Z_\mathrm{[EM:S]}(\mbi{k},\tau)
	-(\rho_\nu+P_\nu)\sigma_\nu
	-(\rho_\gamma+P_\gamma)\sigma_\gamma
\right\}
 \label{eq:phi_psi}\\
\dot{\delta}^\mathrm{(S)}_\mathrm{CDM}
 	&=&
 		-v^\mathrm{(S)}_\mathrm{CDM}+3\dot{\phi}~,\label{eq:CDM_rho}\\
\dot{v}^\mathrm{(S)}_\mathrm{CDM}
 	&=&
        -\frac{\dot{a}}{a}v^\mathrm{(S)}_\mathrm{CDM}+k^2\psi~,\label{eq:CDM_v}\\
\dot{\delta}^\mathrm{(S)}_{\gamma}
 	&=&
 		-\frac{4}{3}v^\mathrm{(S)}_{\gamma}
 		+4\dot{\phi}~,\label{eq:photon_rho}\\
\dot{\delta}^\mathrm{(S)}_{\nu}
 	&=&
 		-\frac{4}{3}v^\mathrm{(S)}_{\nu}
 		+4\dot{\phi}~,\label{eq:nu_rho}\\
\dot{v}^\mathrm{(S)}_{\gamma}
	&=&
		k^2\left(\frac{1}{4}\delta^\mathrm{(S)}_{\gamma}-\sigma_{\gamma}\right)
		+an_e\sigma_T(v^\mathrm{(S)}_\mathrm{b}-v^\mathrm{(S)}_{\gamma})~+k^2\psi,
		\label{eq:photon_v} \\
\dot{v}^\mathrm{(S)}_{\nu}
	&=&
		k^2\left(\frac{1}{4}\delta^\mathrm{(S)}_{\nu}-\sigma_{\nu}\right)+k^2\psi,\label{eq:nu_v} \\
\dot{\delta}^\mathrm{(S)}_\mathrm{b}
	&=&
		-v^\mathrm{(S)}_\mathrm{b}+3\dot{\phi}
 			\label{eq:baryon_rho}  \\
\dot{v}^\mathrm{(S)}_\mathrm{b}
	&=&
			-\frac{\dot{a}}{a}v^\mathrm{(S)}_\mathrm{b}
 			+c^2_sk^2\delta^\mathrm{(S)}_\mathrm{b}
 			+\frac{4\bar{\rho}_\gamma}{3\bar{\rho}_\mathrm{b}}
 			an_e\sigma_T(v^\mathrm{(S)}_{\gamma}-v^\mathrm{(S)}_\mathrm{b})+k^2\psi
\nonumber\\
&&			
			+
			k^2\frac{L_{\mathrm{[EM:S]}}(\mbi{k},\tau)}{\rho_b}~,
%			\frac{3}{4}k^2\frac{L_{\mathrm{[EM:S]}}(\mbi{k},\tau)}{R\rho_\gamma}~,
			\label{eq:baryon_v}
\end{eqnarray} 
where 
$\sigma_{\gamma}$ in the second term on the right-hand side of Eq. (\ref{eq:photon_v}) is the
shear stress of the photons, and
$L_{\mathrm{[EM:S]}}(\mbi{k},\tau)$ is the Lorentz force. 
Here 
\begin{eqnarray}
E^2_\mathrm{[EM:S]}(\mbi{k},\tau) = 
P_\mathrm{[EM:SE]}(k;k_\mathrm{[PMF]},\sigma)\\
L^2_{\mathrm{[EM:S]}}(\mbi{k},\tau) = 
   P_\mathrm{[EM:SE]}(k;k_\mathrm{[PMF]},\sigma)
 + P_\mathrm{[EM:ST]}(k;k_\mathrm{[PMF]},\sigma)
 - P_\mathrm{[EM:SET]}(k;k_\mathrm{[PMF]},\sigma)
\\
Z^2_{\mathrm[EM:S]}(\mbi{k},\tau) = 
   \frac{4}{9}
   P_\mathrm{[EM:SE]}(k;k_\mathrm{[PMF]},\sigma)
 + P_\mathrm{[EM:ST]}(k;k_\mathrm{[PMF]},\sigma)
 - \frac{2}{3}
   P_\mathrm{[EM:SET]}(k;k_\mathrm{[PMF]},\sigma).
\end{eqnarray}
%_/_/_/_/_/_/_/_/_/_/_/_/_/_/_/_/_/_/_/_/_/_/_/_/_/_/_/_/_/_/_/_/_/_/
%_/_/_/_/_/_/_/_/_/_/_/_/_/_/_/_/_/_/_/_/_/_/_/_/_/_/_/_/_/_/_/_/_/_/
%_/_/_/_/_/_/_/_/_/_/_/_/_/_/_/_/_/_/_/_/_/_/_/_/_/_/_/_/_/_/_/_/_/_/
\subsection{Vector Mode}
%_/_/_/_/_/_/_/_/_/_/_/_/_/_/_/_/_/_/_/_/_/_/_/_/_/_/_/_/_/_/_/_/_/_/
We can obtain the evolution of the vector potential $V(\tau, \mathbf{k})$ influenced by a stochastic PMF as \cite{Hu:1997hp,Hu:1997mn}
\begin{eqnarray}
k
\left(
	\dot{V}+2\frac{\dot{a}}{a}V
\right)
=
- 8\pi a^2 G
\left[
2 \Pi_\mathrm{[EM:V]} (\mathbf{k},\tau)
+ p_\gamma\pi_\gamma+p_\nu\pi_\nu
\right]
\label{eq:vectorpoten}
\end{eqnarray}
where the dot denotes a conformal time derivative, while $p_i$ and $\pi_i$ are the pressure and the anisotropic stress of the photons ($i=\gamma$) and neutrinos ($i=\nu$).
Since the vector anisotropic stress of the baryonic plasma fluid is negligible generally,
it is omitted.
Following \cite{Hu:1997hp,Hu:1997mn}, the Euler equations affected by the PMF for the neutrino, photon and baryon velocities, $v^\mathrm{(V)}_\nu$, 
$v^\mathrm{(V)}_\gamma$, and $v^\mathrm{(V)}_b$ are written as
\begin{eqnarray}
\dot{v}_{\nu}^\mathrm{(V)}-\dot{V}
	= -k\left(\frac{\sqrt{3}}{5}\mathcal{Q}^\mathrm{(1)}_{\nu 2}\right)~~,
\label{eq:vneutrino1}\\
 \dot{v}_{\gamma}^\mathrm{(V)}
-\dot{V}
+\dot{\tau}_c
    (
	v^\mathrm{(V)}_{\gamma}-v^\mathrm{(V)}_{b}
	)
=-k\left(
			\frac{\sqrt{3}}{5}\mathcal{Q}^\mathrm{(1)}_{\gamma 2}
   \right)~~,
\label{eq:vphoton1}\\
 \dot{v}_{b}^\mathrm{(V)}
-\dot{V}
+\frac{\dot{a}}{a}(v_{b}^\mathrm{(V)}-V)
-\frac{1}{R}\dot{\tau}_c(v^\mathrm{(V)}_{\gamma}-v^\mathrm{(V)}_{b})\nonumber \\
	=
		k\frac{\Pi_\mathrm{[EM:V]} (\mathbf{k},\tau)}{\rho_b}~~,
\label{eq:vbaryon1}
\end{eqnarray}
where $R\equiv(3/4)(\rho_b/\rho_\gamma)$. Here, the vector Lorentz force is given by 
\begin{eqnarray}
\Pi^2_\mathrm{[EM:V]} (\mathbf{k},\tau) 
= 
P_\mathrm{[EM:V]}(k;k_\mathrm{[PMF]},\sigma)~.
\end{eqnarray}
For the photons 
$v^\mathrm{(V)}_\gamma=\mathcal{Q}^\mathrm{(1)}_1$, while 
$\mathcal{Q}^\mathrm{(1)}_{\nu 2}$ and $\mathcal{Q}^\mathrm{(1)}_{\gamma 2}$ 
are quadrupole moments of the neutrino and photon angular distributions, respectively. These quantities are proportional to the anisotropic stress tensors. Equations (\ref{eq:vneutrino1})-(\ref{eq:vbaryon1}) denote the vector equations of motion for the cosmic fluid, which arise from the conservation of energy-momentum. 
%_/_/_/_/_/_/_/_/_/_/_/_/_/_/_/_/_/_/_/_/_/_/_/_/_/_/_/_/_/_/_/_/_/_/
%_/_/_/_/_/_/_/_/_/_/_/_/_/_/_/_/_/_/_/_/_/_/_/_/_/_/_/_/_/_/_/_/_/_/
%_/_/_/_/_/_/_/_/_/_/_/_/_/_/_/_/_/_/_/_/_/_/_/_/_/_/_/_/_/_/_/_/_/_/
\subsection{Tensor mode}

The tensor mode $\mathcal{H}$ is given by
\cite{Padmanabhan:1993booka,Hu:1997hp,Hu:1997mn,Dodelson:2003booka}
\begin{eqnarray}
\ddot{\mathcal{H}}
+2\frac{\dot{a}}{a}\dot{\mathcal{H}}
+k^2\mathcal{H}=
8\pi G a^2
\left( 
		\Pi_\mathrm{[EM:T]}+
		\Pi^\mathrm{(T)}_\nu
\right)
~~,
\end{eqnarray}
where $\Pi^\mathrm{(T)}_\nu$ is the anisotropic stress for neutrinos.
Here, 
\begin{eqnarray}
\Pi^2_\mathrm{[EM:T]} (\mathbf{k},\tau) 
= 
P_\mathrm{[EM:T]}(k;k_\mathrm{[PMF]},\sigma)~.
\end{eqnarray}
%_/_/_/_/_/_/_/_/_/_/_/_/_/_/_/_/_/_/_/_/_/_/_/_/_/_/_/_/_/_/_/_/_/_/
%_/_/_/_/_/_/_/_/_/_/_/_/_/_/_/_/_/_/_/_/_/_/_/_/_/_/_/_/_/_/_/_/_/_/
%_/_/_/_/_/_/_/_/_/_/_/_/_/_/_/_/_/_/_/_/_/_/_/_/_/_/_/_/_/_/_/_/_/_/
%_/_/_/_/_/_/_/_/_/_/_/_/_/_/_/_/_/_/_/_/_/_/_/_/_/_/_/_/_/_/_/_/_/_/
\subsection{Initial conditions}
%_/_/_/_/_/_/_/_/_/_/_/_/_/_/_/_/_/_/_/_/_/_/_/_/_/_/_/_/_/_/_/_/_/_/
%_/_/_/_/_/_/_/_/_/_/_/_/_/_/_/_/_/_/_/_/_/_/_/_/_/_/_/_/_/_/_/_/_/_/
We adopt the ''compensated Magnetic Modes'' as our initial
conditions \cite{Shaw:2009nf,2010JCAP...08..025C}. 
Since these initial conditions have been  explained
in detail in Appendix B of Ref.\cite{Shaw:2009nf} and Sec.4 Ref.\cite{2010JCAP...08..025C}, and the list of solutions is too long, we give a brief introduction to these in this article as follows. At first a photon fluid is tightly coupled with a baryon fluid in the very early Universe. Therefore we assume the Thomson scattering terms 
 are negligible.
  In this case we can set $k\tau_C \ll 1$ and $\tau_C/\tau \ll 1$. 
 We assume that the baryon has no pressure and the baryon-photon fluid is representable as an ideal fluid and neglect the anisotropic pressure perturbations of the fluid.
%_/_/_/_/_/_/_/_/_/_/_/_/_/_/_/_/_/_/_/_/_/_/_/_/_/
%_/_/_/_/_/_/_/_/_/_/_/_/_/_/_/_/_/_/_/_/_/_/_/_/_/
%_/_/_/_/_/_/_/_/_/_/_/_/_/_/_/_/_/_/_/_/_/_/_/_/_/
\section{\label{sec:results}Results and Discussions}
%_/_/_/_/_/_/_/_/_/_/_/_/_/_/_/_/_/_/_/_/_/_/_/_/_/
%_/_/_/_/_/_/_/_/_/_/_/_/_/_/_/_/_/_/_/_/_/_/_/_/_/
%_/_/_/_/_/_/_/_/_/_/_/_/_/_/_/_/_/_/_/_/_/_/_/_/_/
In this section we will discuss dependencies of
power spectra from the LND-PMF[Eqs.(\ref{p:se}) - (\ref{p:t})] on the
characteristic scale $k_\mathrm{M}$ and variance
$\sigma_\mathrm{M}$, and show results of temperature
and polarization anisotropies of the CMB generated
from the LND-PMF with a modified CAMB code \cite{camb}.  We will, also,
discuss how the strength of the LND-PMF is constrained by the CMB
observations.
%_/_/_/_/_/_/_/_/_/_/_/_/_/_/_/_/_/_/_/_/_/_/_/_/_/
%_/_/_/_/_/_/_/_/_/_/_/_/_/_/_/_/_/_/_/_/_/_/_/_/_/
\subsection{Effects of the LND-PMF on the CMB \label{subsec:k_and_sigma_dependence}}
%_/_/_/_/_/_/_/_/_/_/_/_/_/_/_/_/_/_/_/_/_/_/_/_/_/
%_/_/_/_/_/_/_/_/_/_/_/_/_/_/_/_/_/_/_/_/_/_/_/_/_/
Figures \ref{fig2} and \ref{fig3} show the four
modes of the CMB (TT, EE, BB, and TE modes) from the LND-PMF.  Since
qualitative features of the scalar, vector, and tensor modes on the CMB
are almost the same as previous studies \cite{Kojima:2008rf,
2008PhRvD..78f3012K, Yamazaki:2009na, 2010PhRvD..81j3519Y}, in this
article, we will briefly explain these features of
the CMB with the PMF as follows.  

In general, the vector mode dominates all the temperature and
polarization auto- and cross-correlation anisotropies at $\ell \gtrsim
500$, while the scalar mode dominates at smaller $\ell$, except for the
BB mode.  For BB mode of the CMB, the vector mode dominates the power
spectrum for all $\ell$.

To understand the feature of temperature fluctuations and polarization
anisotropies of the CMB from the LND-PMF in Fig.\ref{fig2}, we shall go
back to the definitions $\mathrm{Eqs.}(\ref{eq:power_lg}), (\ref{p:se})
- (\ref{p:t})$ and show Fig.\ref{fig1} which depicts $P(k)$ and $P(k)
\times k^3$.  From the definition of Eq.(\ref{eq:power_lg}), we can
understand that the peak amplitudes of the LND-PMF power spectra of each
$ k_\mathrm{M}$ for the CMB spectra are along the scale invariant
amplitude [see panel (c) of Fig.\ref{fig1}].  Therefore, the
superposition of the spectra of CMB anisotropy from different $
k_\mathrm{M}$ corresponds to the spectra from the scale invariant
magnetic fields.  Indeed in the top-center panel of
Fig.\ref{fig2}, we find that the points of the CMB power spectra on the same
$\sigma_\mathrm{M}$, which are corresponding to the peak of the
$P(k)\times k^3$, are distributed along the scale invariant curve.
We should note that the peak position of the LND-PMF
source $P(k) \times k^3$ is different from one of the eventual power
spectra of the CMB, because the final spectrum should be obtained by
multiplying the LND-PMF power spectrum by
the linear transfer functions which include various physical effects
including the PMF.

Considering the mathematical definition of Eq.(\ref{eq:power_lg})
as mentioned above, 
the peak scale of the LND-PMF power spectrum is determined by
$k_\mathrm{M}$. We find, however, that the peak positions of the CMB
spectra from the
PMF power spectrum are characterized not only by $k_\mathrm{M}$ but also
by $\sigma_\mathrm{M}$.  The reason is as follows.
The LND-PMF power spectrum $P(k)$
becomes more broad with increasing
$\sigma_\mathrm{M}$ as shown in panel (b) of Fig.3. 
{The amplitudes of the CMB 
power spectra are determined by $k^3 P(k)$ 
which is plotted in the panel (d) of Fig.3.
The peak position in terms of the combination $k^3 P(k)$ is 
affected also by $\sigma_\mathrm{M}$, and hence the positions of the
peaks in CMB power spectra are also affected by $\sigma_\mathrm{M}$.

Considering the scale dependencies of the scalar and vector modes
\cite{Paoletti:2008ck,Finelli:2008xh,Shaw:2009nf,2010PhRvD..81b3008Y},
in the case of the smaller $k_\mathrm{M}$ and/or smaller 
$\sigma_\mathrm{M}$ the scalar mode determines the maximum amplitude
of the power spectra of the CMB and dominates the
power spectra at smaller angular moments $\ell < 1000$.  
Meanwhile,
in the case of the larger $k_\mathrm{M}$ and/or larger $\sigma_\mathrm{M}$, 
the vector mode determines the maximum amplitude
and dominates at larger $\ell$.
In this case, since the peak of the total power
spectrum from the LND-PMF lies among 1000 $<\ell<$ 2000 (e.g., for
$\sigma_\mathrm{M} = 1.0$, $k_\mathrm{[LND]} \sim 1.0^{-2}$, see Fig.4), 
we can expect that the observation data of the CMB among 1000 $<\ell<$
2000 better constrain LND-PMF parameters.
%_/_/_/_/_/_/_/_/_/_/_/_/_/_/_/_/_/_/_/_/_/_/_/_/_/
%_/_/_/_/_/_/_/_/_/_/_/_/_/_/_/_/_/_/_/_/_/_/_/_/_/
\subsection{How are LND-PMF parameters constrained by the CMB?}
%_/_/_/_/_/_/_/_/_/_/_/_/_/_/_/_/_/_/_/_/_/_/_/_/_/
%_/_/_/_/_/_/_/_/_/_/_/_/_/_/_/_/_/_/_/_/_/_/_/_/_/
In the above sections, we have explained the behaviors of the power
spectrum of the PMF from LND distributions and the effects on the CMB.
In this subsection, considering these features, we discuss how LND-PMF
parameters, $k_\mathrm{M}$, $\sigma_\mathrm{M}$ and
$B_\mathrm{LND}$ are constrained by the CMB observations.  As mentioned
in the above sections, the peak positions of the spectra of PMF
electromagnetic tensors mainly depend on the characteristic wavenumber
$k_\mathrm{M}$ and $\sigma_\mathrm{M}$, the amplitudes mainly
depend on $k_\mathrm{M}$ and $B_\mathrm{LND}$, and the features
the peak width and the damping scale depend on the
$\sigma_\mathrm{M}$.  Figures \ref{fig4} and \ref{fig5} show the CMB
(TT, EE, BB and TE) angular power spectra from primary (standard
adiabatic mode without the PMF) and from 
the LND-PMF along with observations
(WMAP \cite{2011ApJS..192...16L}, ACBAR \cite{2009ApJ...694.1200R}, CBI \cite{Sievers:2005gj}
and QUaD \cite{2009ApJ...705..978B}).  Considering the above mentioned
facts, 
if $k_\mathrm{M}$
or $\sigma_\mathrm{M}$ is 
larger,
the PMF amplitude is constrained by the CMB observation data at
smaller scales. On the other hand, 
if $k_\mathrm{M}$
or $\sigma_\mathrm{M}$ is
smaller,
the PMF amplitude is constrained by the
CMB at larger scales.  The strength of the PMF $B_\mathrm{LND}$
monotonically increases or decreases the amplitude of CMB power spectra
from the LND-PMF.  These features may suggest a significant
anticorrelation between the characteristic wavenumber
$k_\mathrm{M}$ and the variance $\sigma_\mathrm{M}$. Meanwhile,
the slopes of the spectrum at small scale are mainly characterized
 by the variance $\sigma_\mathrm{M}$.

To research how the strength of the LND-PMF is constrained by the CMB
observations on each characteristic scale $k_\mathrm{M}$, 
we perform a Markov chain Monte Carlo analysis with the CMB observations
and obtain the constraint on the $B_\mathrm{LND}$-$k_\mathrm{M}$
plane.  Since our purpose is qualitatively understanding 
how the primordial magnetic fields are constrained scale by scale
the variance parameter of the LND-PMF is fixed at
$\sigma_\mathrm{M} = 1$ for simplicity and we also fix the
cosmological parameters to those of the best-fit flat $\Lambda$CDM model
as given in the WMAP 7yr analysis \cite{2011ApJS..192...16L}.

Figure \ref{fig6} shows the constrained strengths of the LND-PMF from the CMB observations (WMAP 7th, ACBAR, CBI, Boomerang, and QuaD) for each scale. From this figure, the LND-PMFs at $10^{-3}$ Mpc $ \le~k~\le~ 10^{-2}$ Mpc are most tightly constrained. 
%_/_/_/_/_/_/_/_/_/_/_/_/_/_/_/_/_/_/_/_/_/_/_/_/_/
%_/_/_/_/_/_/_/_/_/_/_/_/_/_/_/_/_/_/_/_/_/_/_/_/_/
%_/_/_/_/_/_/_/_/_/_/_/_/_/_/_/_/_/_/_/_/_/_/_/_/_/
\section{Summary}
%_/_/_/_/_/_/_/_/_/_/_/_/_/_/_/_/_/_/_/_/_/_/_/_/_/
%_/_/_/_/_/_/_/_/_/_/_/_/_/_/_/_/_/_/_/_/_/_/_/_/_/
%_/_/_/_/_/_/_/_/_/_/_/_/_/_/_/_/_/_/_/_/_/_/_/_/_/
It is natural to assume that the power spectrum of the PMF from
inflation is the power law.  However, when we consider PMF generated
by causal mechanisms, it would be appropriate to assume a spectrum with
a characteristic scale. In this paper, we assumed the log-normal
distribution as the PMF spectrum and studied the effects of the PMFs
on the temperature and polarization anisotropies of the CMB.  We have
revealed the features of the LND-PMF effects on the CMB as follows: (1)
For larger $k_\mathrm{M}$ and/or
$\sigma_\mathrm{M}$, 
the CMB spectra of TT, TE and EE from the LND-PMF are dominated by the
vector mode. Meanwhile, in the opposite case, these spectra
are dominated by the scalar mode.  (2) The CMB spectrum of the BB mode
for all scales and other modes for smaller scales is dominated by
the vector modes. Because three parameters
which characterize the LND-PMF affect the CMB power spectra differently
at small and large scales we expect that
tight constraints can be placed on these parameters without degeneracy.
We found that the LND-PMF which generates CMB anisotropies among 1000
$<\ell<$ 2000 is most effectively constrained by the current CMB data
sets.
 For example, $B_\mathrm{LND}$ for
$10^{-3}\mathrm{Mpc}^{-1}~\le~k/h~\le 10^{-2}\mathrm{Mpc}^{-1}$ and
$\sigma_\mathrm{M} = 1.0$ is limited most strongly as shown in
Fig.\ref{fig6}.  
In the near future, the tighter
constraints on $B_\mathrm{LND}$ at $k_\mathrm{M}/h>10^{-2}$Mpc$^{-1}$ will be expected
from the observations, such as the Planck, QUIET,
and PolarBear missions.
\begin{figure}
\includegraphics[width=1.0\textwidth]{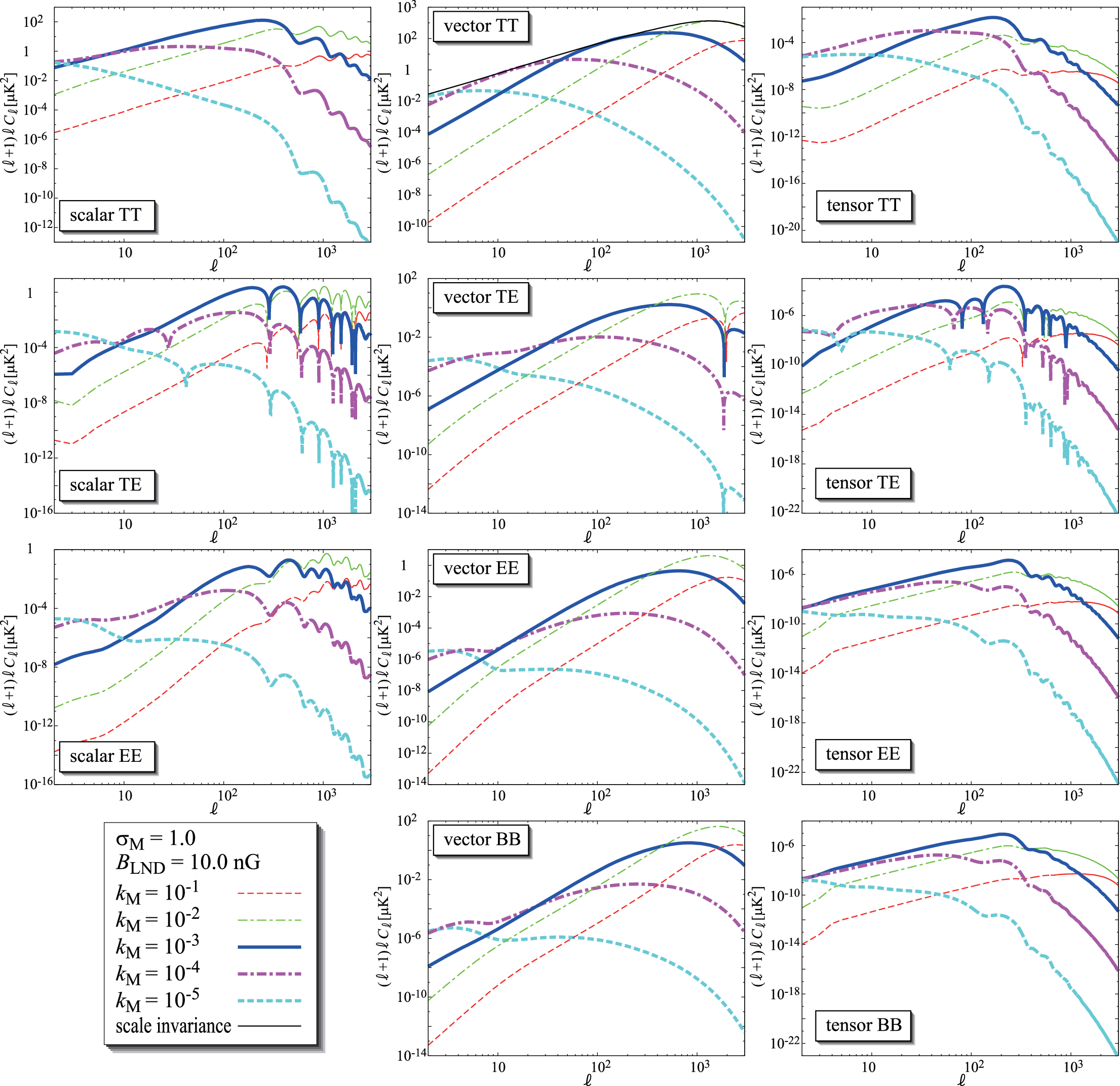}% Here is how to import EPS art
\caption{\label{fig2}
CMB spectra from the LND-PMF at $\sigma_\mathrm{M}=1.0$ and $B_\mathrm{LND} = $10nG.
TT, EE, BB and TE in each panel indicate the temperature auto-correlation, the E-mode auto-correlation, the B-mode auto-correlation and the temperature E-mode cross correlation, respectively
Curves in all panels are theoretical lines as indicated in the legend box at the left bottom.
Curves of all TE modes are plotted in the absolute value.
} 
\end{figure}

\begin{figure}
\includegraphics[width=1.0\textwidth]{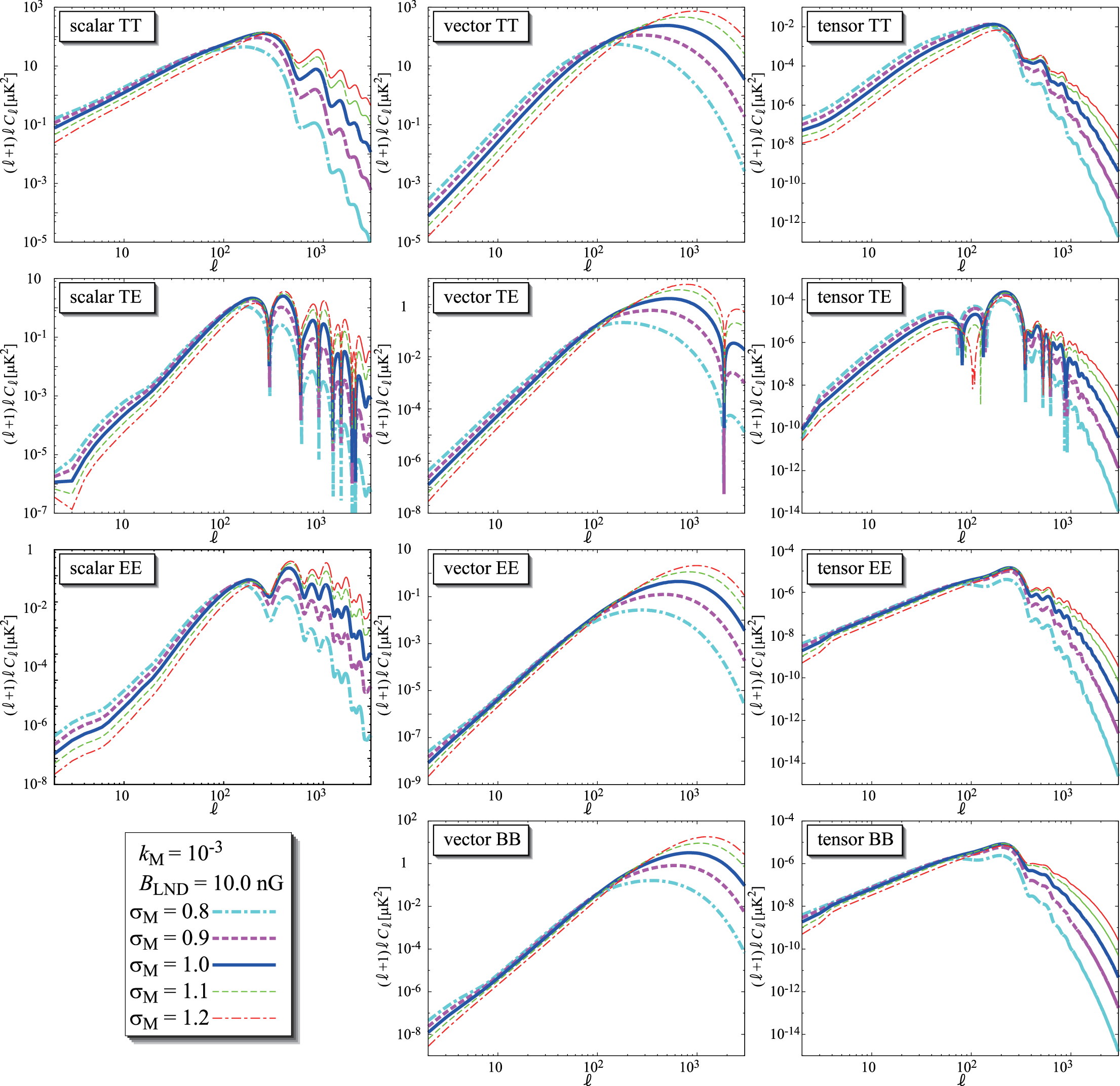}% Here is how to import EPS art
\caption{\label{fig3}
CMB spectra from the LND-PMF at $k_\mathrm{M}=10^{-3}$ and $B_\mathrm{LND} = $10nG.
Curves in all panels are theoretical lines as indicated in the legend box at the left bottom.
Curves of all TE modes are plotted in the absolute value.
} 
\end{figure}

\begin{figure}
\includegraphics[width=0.8\textwidth]{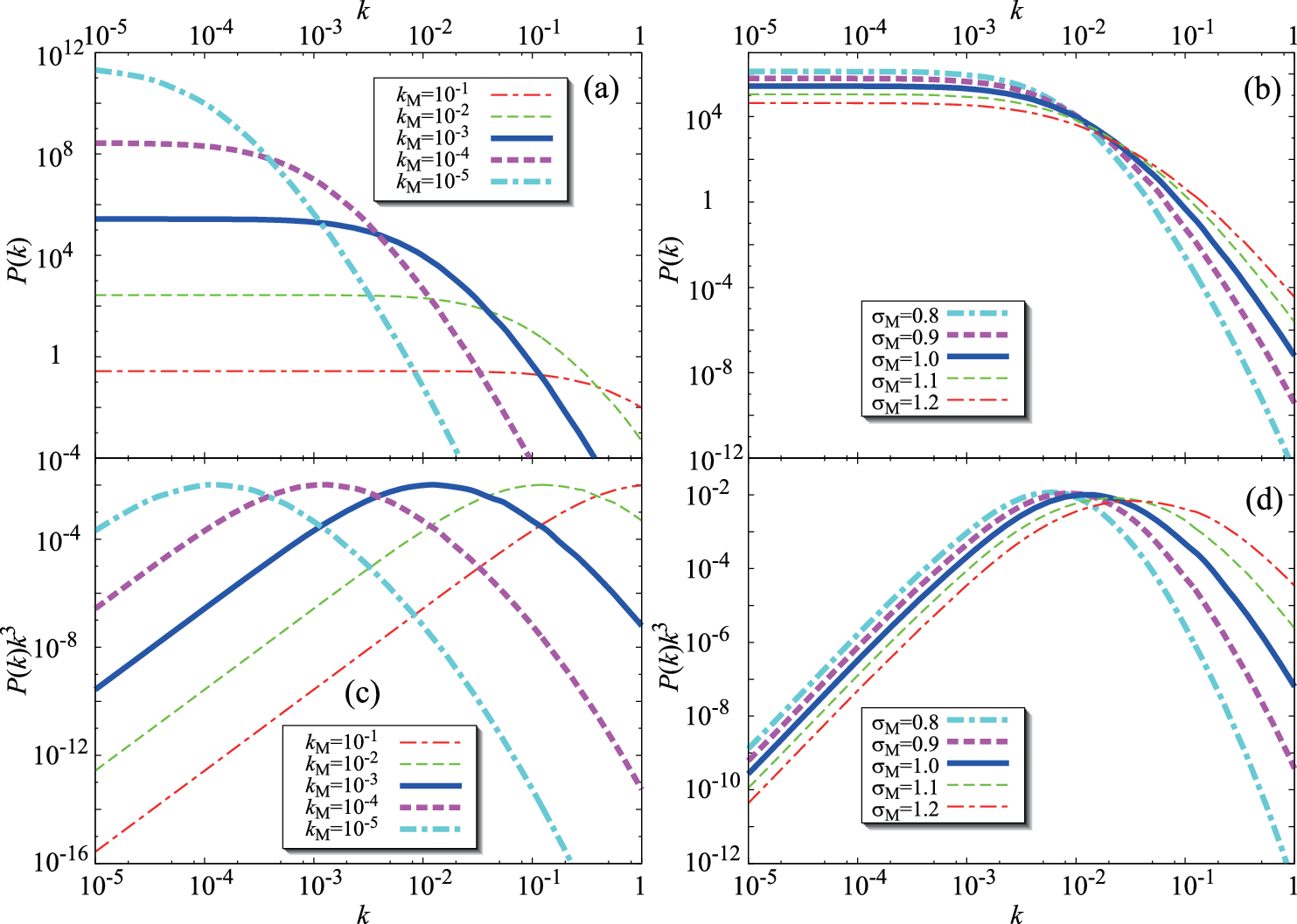}% Here is how to import EPS art
\caption{\label{fig1}
Power spectrum of the energy density of the PMF. 
Panels (a) and (c) on this figure are for the fixed $\sigma_\mathrm{M}$ value at $\sigma_\mathrm{M}=1.0$, 
and panels (b) and (d) are for the fixed $k_\mathrm{M}$ value at $k_\mathrm{M}=10^{-3}$.  
Curves in different colors in all panels correspond to the different PMF parameters as indicated in the legend box in the figure. 
Note that from Eqs.(\ref{p:se})-(\ref{p:t}) we see that the other spectra, 
such as magnetic tension, anisotropic stress and so on, 
have very similar shape to that of the energy density so we only show the spectrum of the energy density in the figure.} 
\end{figure}

\begin{figure}
\includegraphics[width=1.0\textwidth]{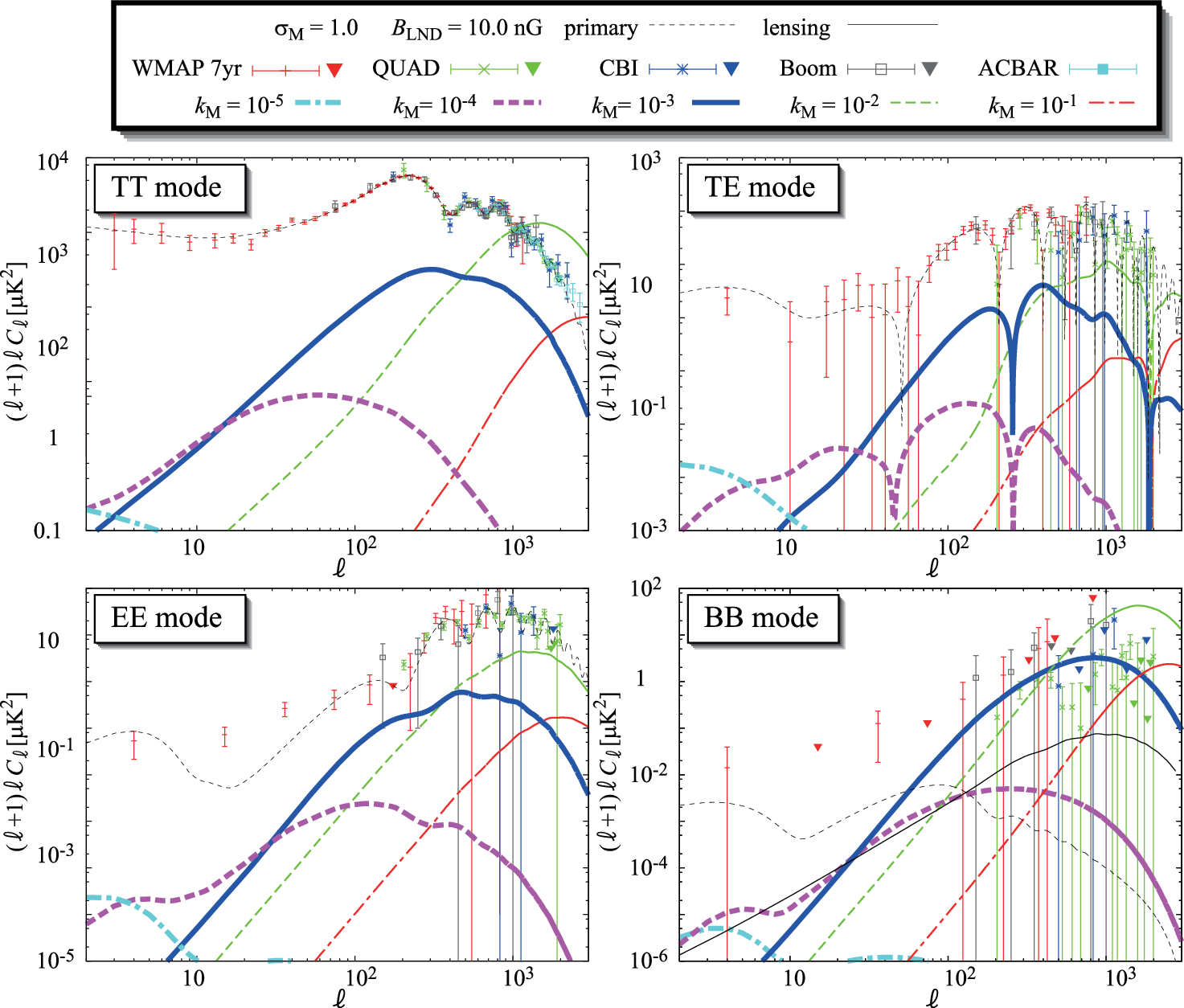}% Here is how to import EPS art
\caption{\label{fig4}
Comparison of total CMB spectra from the LND-PMF at $\sigma_\mathrm{M}=1.0$ and $B_\mathrm{LND} = $10nG with the best-fitting CMB power spectra without the LND-PMF (dashed line).
Curves in all panels are the theoretical lines as indicated in the legend box on the figure.
Curves of the TE mode are plotted in the absolute value.
} 
\end{figure}

\begin{figure}
\includegraphics[width=1.0\textwidth]{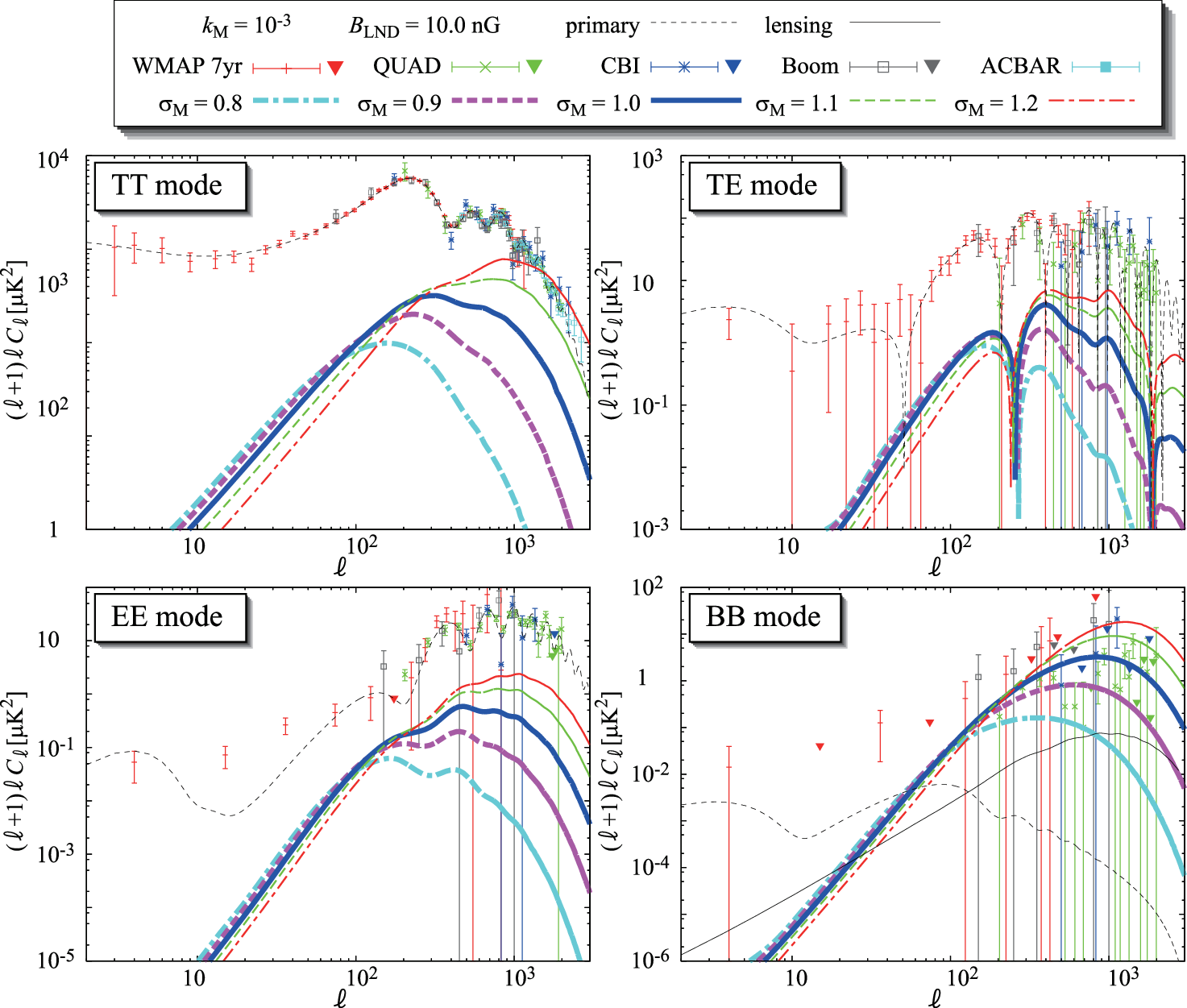}% Here is how to import EPS art
\caption{\label{fig5}
Comparison of total CMB spectra from the LND-PMF at $k_\mathrm{M}=10^{-3}$ and $B_\mathrm{LND} = $10nG with the best-fitting CMB power spectra without the LND-PMF (dashed line).
Curves in all panels are the theoretical lines as indicated in the legend box on the figure.
Curves of the TE mode are plotted in the absolute value.
} 
\end{figure}

\begin{figure}
\includegraphics[width=1.0\textwidth]{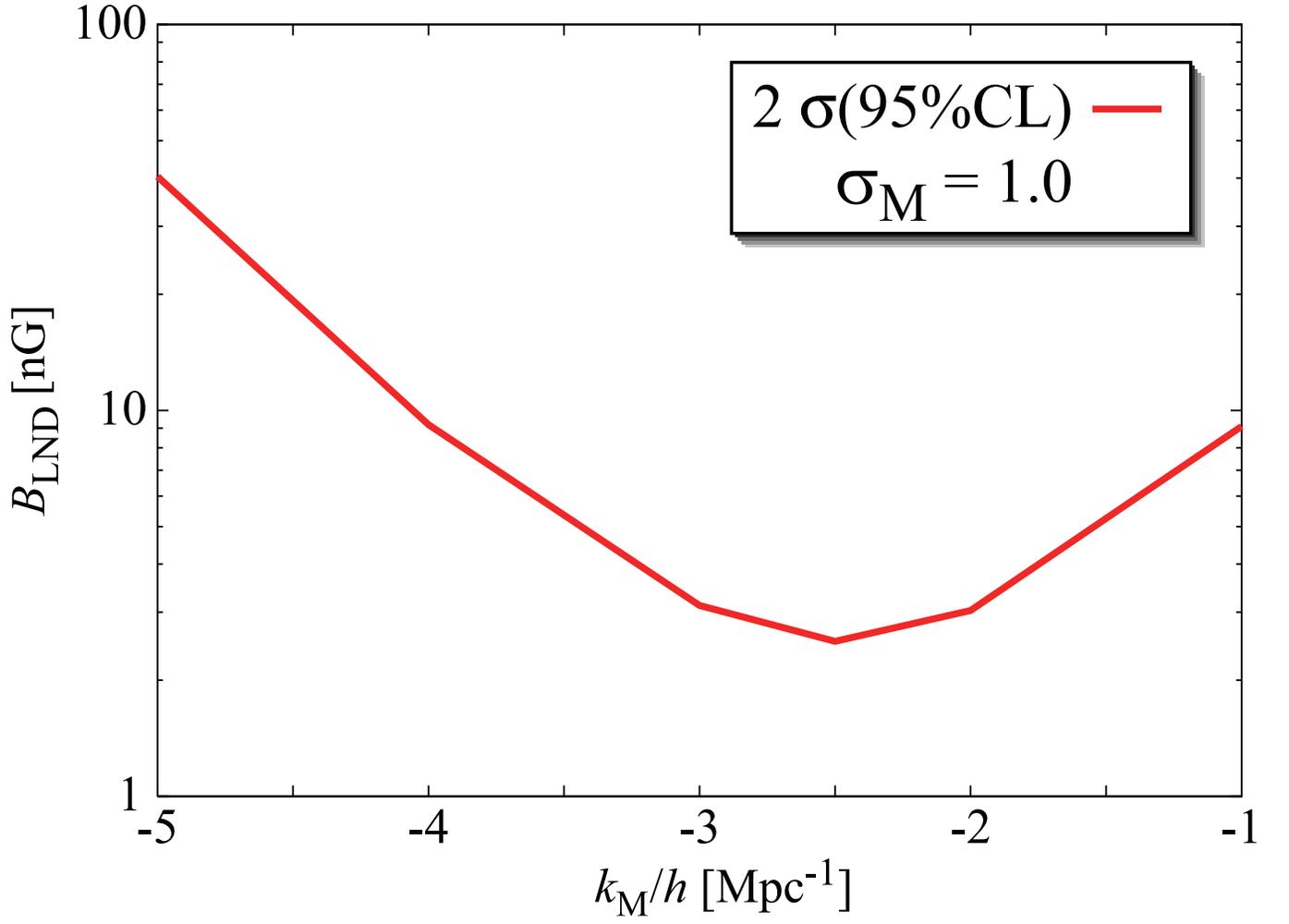}% Here is how to import EPS art
\caption{\label{fig6}
Constraint on the strengths of LND-PMF for $10^{-5}< k_\mathrm{M} <10^{-1}$. The bold curve is the 2 $\sigma$ upper limits of $B_\mathrm{LND}$ [nG]. We fix the standard cosmological parameters and use the best-fitted value from WMAP 7th + tensor mode\cite{2011ApJS..192...16L}.
} 
\end{figure}
\begin{acknowledgments}
This work has been supported in part by Grants-in-Aid for Scientific Research 
(K.I.,21740177 and 22012004, and K. T.,21840028) 
of the Ministry of Education, Culture, Sports,
Science and Technology of Japan.
\end{acknowledgments}
\bibliographystyle{apsrev}

\end{document}